\documentclass{icrc29}
\usepackage{graphicx,amssymb,amsmath,times}
\begin{document}
%Title of paper
\title[Primary composition using asymmetries ...]{Hinting at primary composition using asymmetries in time distributions}
\author[M. T. Dova et al.] {Dova, M. T.$^a$, Mancenido, M $^a$, Mariazzi, A.G. $^a$ $^c$, Arqueros, F.$^b$, Garcia-Pinto, D.$^b$
\newauthor
\\
        (a) Dpto.de Fisica,Universidad Nacional de La Plata, C.C.67, 1900 La Plata,Argentina.
\\
        (b) Facultad de Ciencias Fisicas, Universidad Complutense, E-28040 Madrid, Spain
\\	(c) Now at School of Physics and Astronomy, University of Leeds,Leeds LS2 9JT,UK
        }
\presenter{Presenter: Dova, Maria-Teresa (dova@fisica.unlp.edu.ar), \  
arg-dova-MT-abs1-he14-oral}

\maketitle

\begin{abstract}

Evidence of azimuthal asymmetries in the time structure and signal 
size have been found in non-vertical showers at the Pierre Auger 
Observatory. It has been previously shown that the asymmetry in 
time distributions offers a new possibility for the determination 
of the mass composition. New studies have demonstrated that 
the dependence of the asymmetry parameter in the rise-time and 
fall-time distributions with $sec\theta$ shows a clear peak. 
Both, the position of the peak, $X_{asymax}$, and the size of the 
asymmetry at $X_{asymax}$ are sensitive to primary mass composition 
and have a small dependence on energy. In this paper a study 
of the discriminating power of the new observables to separate 
primary species is presented.
\end{abstract}

\section{Introduction}
As it is well known, the circular symmetry observed in the signals
collected by the surface detectors
is broken in the case of inclined showers.
Evidence of the azimuthal asymmetries in the signal size were first observed 
at Haverah Park ~\cite{Engl} and the first observations of asymmetries in the
time structure of the ground detector signals were found in the 
Pierre Auger Observatory~\cite{Dova:2003rz}. 
The azimuthal dependence arises mainly due to the different
paths traveled by the particles at different azimuth angles.
%Let's first define the coordinates we will use in this work.
%The incoming direction of the inclined shower corresponds to 
%the origin of 
%the azimuth angle $\zeta=0$ in the shower plane.  
%The regions near $\zeta=0$ and $\zeta=\pi$ are called 
%``early'' and  ``late'' regions respectively,  because the shower is in a different
%stage of development at each side of the core.
%Then, it is clear that in a very inclined shower 
%one expects an azimuthal dependence, due to the different
%paths traveled by the particles at different azimuth angles.
In ground array experiments the analysis is done projecting the collected
signals and time distributions at ground level into the shower plane
neglecting the shower evolution. 
%If  the radial symmetry around the shower axis were maintained 
%in an inclined shower, then the 
%density would be a function of the core distance in the shower plane and 
%the slant depth along the shower axis.
This results in an azimuth angle
dependence in the slant depth,
as it was proposed in ~\cite{Dova:2001jy}.\\
%There is another effect related with the shape of the detectors, the resulting asymmetry being small
% at the angles considered here ~\cite{Pierre, daSilva}.
The observed asymmetry in the time distributions offers a new possibility
for primary composition determination, because its magnitude is
strongly dependent on the muon to electromagnetic ratio at ground.
We have done a preliminary 
study on the information in time distributions
using simulations of proton and iron initiated showers
with the aim of estimating the sensitivity of the Pierre Auger Observatory
for hadronic primary discrimination~\cite{Dova:2003rz}. 
The following observables were analyzed:``rise-time'' (defined as the time between the 10\% and 50\%  
of the integrated signal) and ``fall-time'' (time between the 50\% 
and 90\%  of the integrated signal).
These variables were studied as a function of
the azimuth angle in the shower plane at
fixed core ranges and zenith angles, for showers initiated by proton and iron primaries.\\
In this paper we describe a new observable related 
to the asymmetry factor of the time structure of the signal 
in the water-\v{C}erenkov detectors of the Pierre Auger Observatory 
useful to discriminate composition.

\section{Asymmetries as an indication of shower evolution and mass composition}
Most of the observables related to 
composition are like a snapshot of the shower development 
and then, are correlated with $X_{max}$ and the atmospheric depth.
The time distribution of the signals contains implicitly 
the information of the shower development.
Therefore, it is natural to expect a dependence of the mean values of ``rise-time'' 
and ``fall-time'', and the corresponding asymmetries observed, with the atmospheric depth.\\
%This dependence with depth is reflected in the variation
%with zenith angle or atmospheric pressure.
%In this work we studied the dependence with zenith angle,
%in particular as a function of $sec\theta$, due to the fact that
%the slant depth is proportional to $sec\theta$ with a factor $t$, 
%where $t$ is depth of the experiment.
%%%%%%%%%%%%%%%%%%%%%%%%%%%%%%%%%%%%%%%%%%%%%%%%%%%%%%%%%%5
%The sensitivity of timing parameters to primary composition 
%can be explained on the basis of the dominance 
%in the different portions of the signal
%of the electromagnetic or muonic component.
%The first portion of the signal is dominated by
%the muon component that tends to arrive earlier and over a short period of time,
%while the electromagnetic particles are spread out on time.
%The limit between these two portions of the signal
%depends on the shower core distance.
%The observed azimuthal asymmetry in the timing parameters
%is also an indicator of composition because it is directly related
%to the gradual absorption of the electromagnetic component
%in the ``late'' region with respect to the ``early'' region in very
%inclined showers, changing the ratio of the muonic to electromagnetic components.
%There is a dependence of 
%the atmospheric slant depth with the azimuth angle $\zeta$  for inclined showers
%as it was proposed in ~\cite{Dova:2001jy}.
%The regions around 
%$\zeta=0$ and $\zeta=\pi$ are called ``early'' and ``late''respectively.
%%%%%%%%%%%%%%%%%%%%%%%%%%%%%%%%%%%%%%%%%%%%%%%%%%%%%%%%%%%%
If we call $t'$ the atmospheric slant depth, 
$t'=\int_h^\infty \rho_{atm}(z) \,\,dz'$ 
with $z'$ along the shower axis, then, $t'(\zeta)  = 
t\,\sec\theta\,(1+ B_0\,\tan\theta \,\cos\zeta) = t_s\,+\,\Delta
t_s(\zeta)$. We can treat this dependence in a first approximation using a Taylor 
expansion in slant depth around $t\,\sec\theta$ keeping only the first term,
considering that the azimuth angle correction is small compared to the
slant depth. 
This is equivalent to using a linear function in cosine of the 
azimuth angle  in the shower plane, $\zeta$ (defined with
$\zeta=0$ in the incoming direction of the shower) to describe asymmetries in data. \\
In vertical showers, a generic time distribution $\tau(r,t)$ will be
function of $t$, the atmospheric depth along the shower axis,
and $r$, the core distance perpendicular to the shower axis. For
inclined showers, $\tau(r,t)\, \,\rightarrow\, \,\tau(r,t'(\zeta,\theta))$.
%\begin{eqnarray}
%t'(\zeta) & = & t\,\sec\theta\,(1+B\,\cos\zeta) = t_s\,+\,\Delta t_s(\zeta)\\
%B & = & +B\,\cos\zeta) = t_s\,+\,\Delta t_s(\zeta)$
%\begin{eqnarray}
%t'(\zeta) & = & t\,\sec\theta\,(1+B\,\cos\zeta) = t_s\,+\,\Delta t_s(\zeta)\\
%B & = & B_0\,\tan\theta \\
%t_s & = & t\,\sec\theta\,\\
%\Delta t_s(\zeta) & = & t_s B\,\cos\zeta\\
%\end{eqnarray}
%%%%%%%%%%%%%%%%%%%%%%%%%%%%%%%%%%%%%%%%%%%%%%%%%%%%%%%%%%%
\begin{figure}[t]
\begin{minipage}[t]{7.5cm}
\begin{center}
\includegraphics*[width=1.0\textwidth,angle=0,clip]{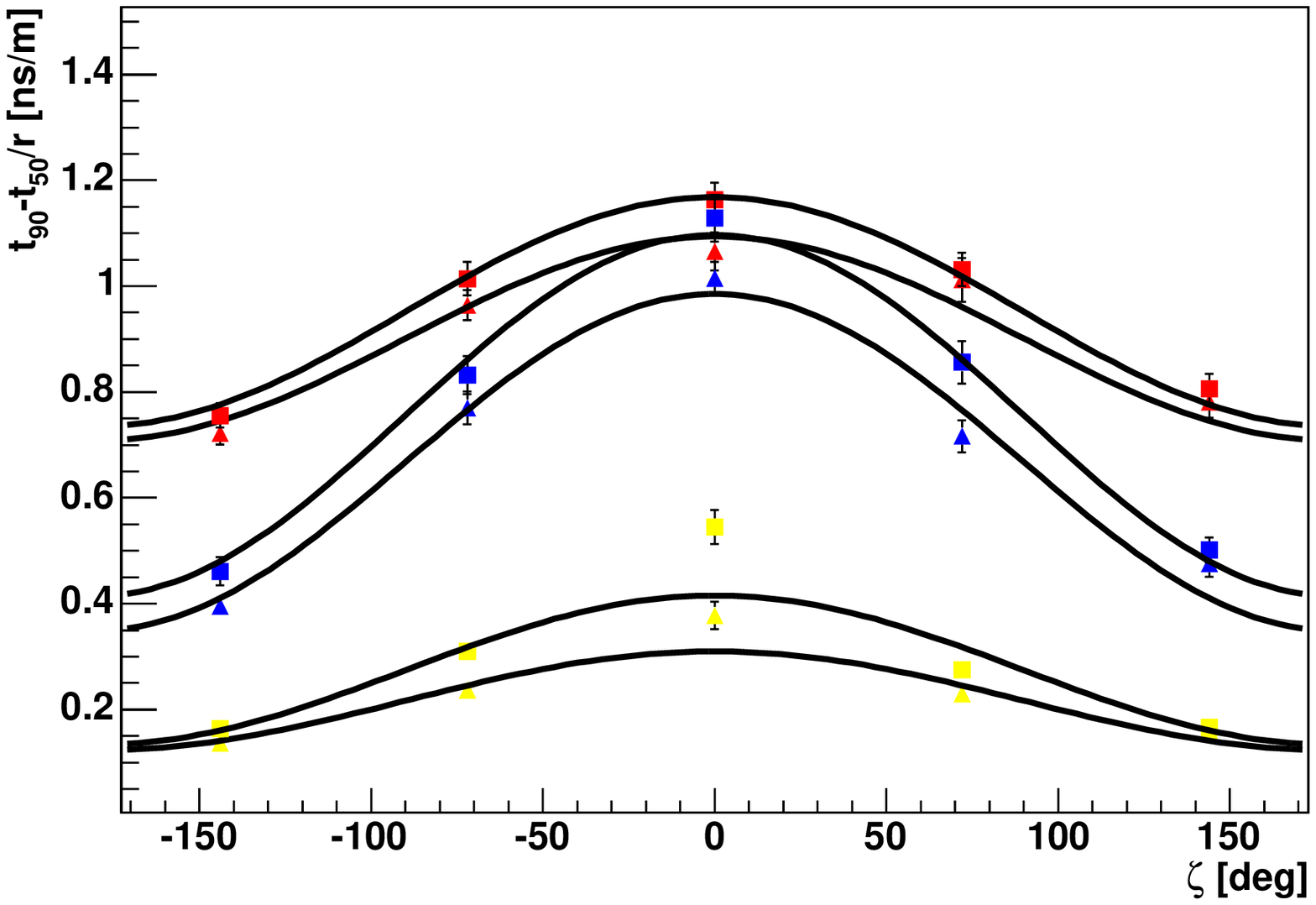}
\end{center}
\caption{Fall-time as a function of $\zeta$ for 100 EeV for proton ({\tiny $\square$}) and iron ($\blacktriangle$) for 25, 45 and 60 from top to bottom.}
\label{fall-zeta}
\end{minipage}
\hfill
\begin{minipage}[t]{7.5cm}
\begin{center}
\includegraphics*[width=1.0\textwidth,angle=0,clip]{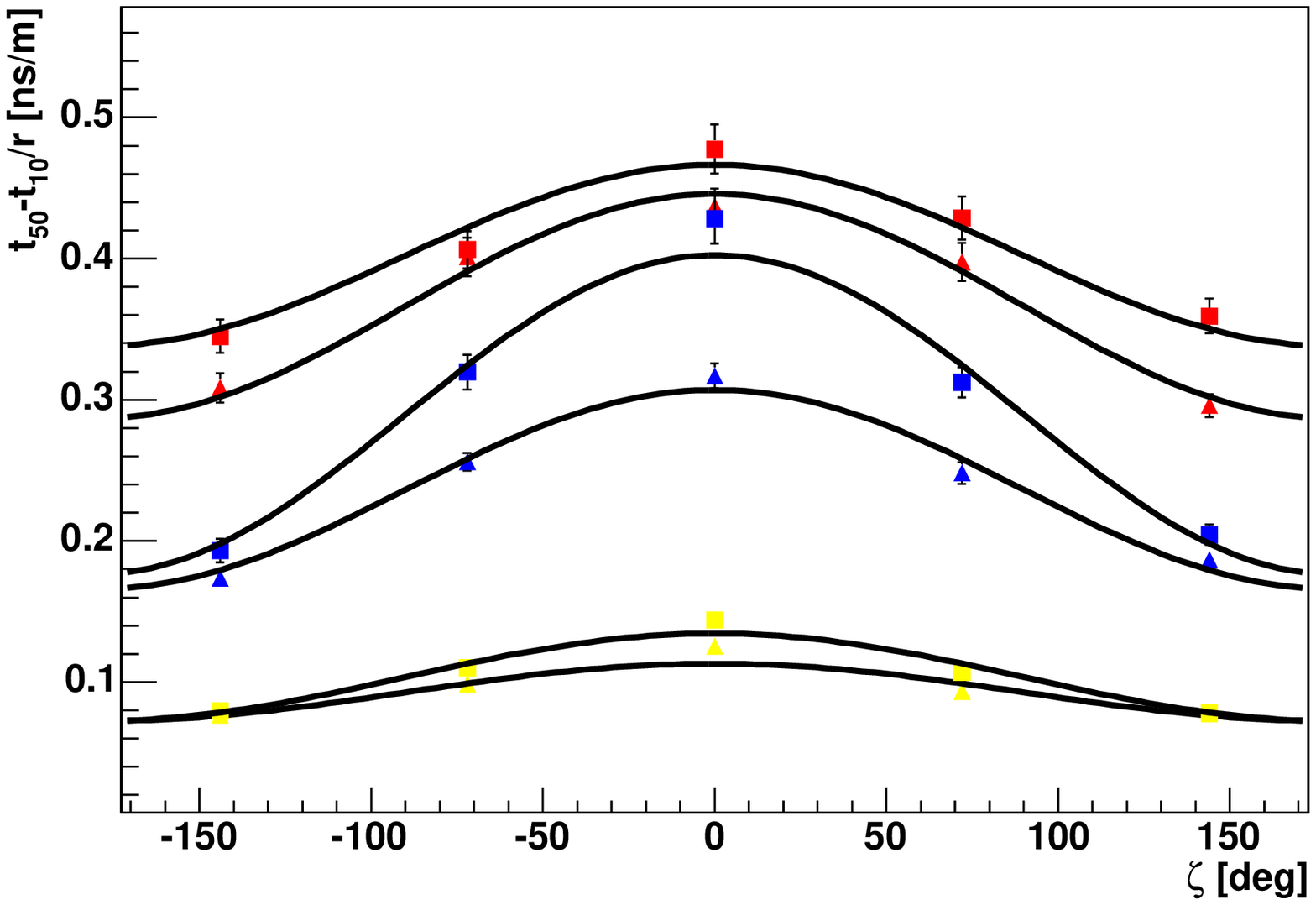}
\end{center}
\caption{Rise-time as a function of $\zeta$ for 100 EeV for proton ({\tiny $\square$}) and iron ($\blacktriangle$) for 25, 45 and 60 from top to bottom}
\label{rise-zeta}
\end{minipage}
\hfill
\end{figure}
%\begin{eqnarray}
%\tau(r,t)\, \,\rightarrow\, \,\tau(r,t'(\zeta,\theta))
%\end{eqnarray}
where $t'$ is the atmospheric depth along the shower axis
and $r$ the core distance in the shower plane.
Performing a Taylor expansion around $t_s$, we obtain:
\begin{eqnarray}
%\tau(r,\zeta) & = & \tau(r,t'(\zeta))=\tau(r,t_s+\Delta t_s(\zeta))\\
%\tau(r,\zeta) & = & \tau(r,t_s)+\frac{\partial\tau}{\partial t'}|_{t_s}\,\Delta t_s(\zeta))+...\\ 
%\tau(r,\zeta) & = & \tau(r,t_s)+\frac{\partial\tau}{\partial t'}|_{t_s}\,\,t_s\,B\,\cos\zeta+...\\
\tau(r,\zeta) = \tau(r,t_s)(1+\frac{\partial\log\tau}{\partial log t'}|_{t_s}\, B\,\cos\zeta+...)
\end{eqnarray}
If $\tau(r,\zeta) = a + b\,\cos\zeta$, the asymmetry factor $\frac{b}{a}$
depends on the core distance and the atmospheric depth,
and it is a measurement of the logarithmic rate of change
of the variable considered.

% $\tau$ with atmospheric depth:
%\begin{eqnarray}
%\tau(r,\zeta) & = & a + b\,\cos\zeta\\
%a & = & \tau(r,t\sec\theta)\\
%\frac{b}{a} & = & B\,\frac{\partial\ln\tau}{\partial\ln t'}|_{t_s}
%\end{eqnarray}
%%%%%%%%%%%%%%%%%%%%%%%%%%%%%%%%%%%%%%%%%%%%%%%%%%%%%%%%%%5
%Then, the asymmetry factor $\frac{b}{a}$ depends as previously stated experimentally, on  $t_s$ and as a consequence on the longitudinal development of the shower in the atmosphere.
%As this is an indicator of shower evolution,
%we expect it to be related to the composition of the primary particle.
Studying asymmetries in time distributions from this point of view 
allowed us to find a new observable in the dependence of the asymmetry factor 
with depth, useful to discriminate primary composition.\\
%\section{Monte Carlo studies}
For the present Monte Carlo study, we use proton and iron initiated 
showers generated with {\sc aires} 2.6.0 /{\sc qgsjet01} 
with primary energies $10^{19}$ eV and $10^{20}$ eV and zenith angles between  $0^0$ and $60^0$ ($sec\theta=$ 1.103, 1.221, 1.414, 1.743 and 2).
The detector response of the Pierre Auger Observatory was simulated (using the official simulation and reconstruction tool, Offline v1r2) with the SDSim module         
in the full array configuration and with all the parameters in their default values. The resulting 
data were reconstructed using the reconstruction modules. The sample used consisted in 20 showers for $10^{19}$ eV protons, 25 showers for $10^{19}$ eV irons and 30 showers for $10^{20}$ eV proton and iron each, for each angle considered.

%The showers were thrown 20 times, using 
%SDSim/v2r8(AugerSoftRelease/3.00). 
%We studied the variation of ``rise-time'' and ``fall-time'' with
%atmospheric depth and primary energy.
The dependence of the ``rise-time'' and ``fall-time'' with azimuth angle was fitted 
using the functional dependence, $\tau_{}= a + b cos\zeta$. In Figures
~\ref{fall-zeta} and ~\ref{rise-zeta} we show the mean value of the timing distributions divided by core
distance for proton and iron primaries as a function of $\zeta$ for
the case of 
primary energy $10^{20}$eV.
%\begin{equation}
%\tau_{}= a + b cos\zeta 
%\end{equation}
%where $a$ represents the mean value and $b/a$ is the asymmetry factor.
The fit was performed for each primary species, energy and zenith angle, for all stations between 
 $500m < R < 2000m$ from the core.
It is important to point out that the asymmetry factor is different 
for different primaries, as one expects due to the greater number of muons 
in showers initiated by heavy nuclei than in showers initiated by protons.\\
We studied then the behavior of the asymmetry factor with depth
represented by $\sec\theta$.
The corresponding distributions show a clear peak in atmospheric depth,
which is in a different position for proton and iron showers.
Two examples of these distributions are presented in Figures 3 and 4,
corresponding to the asymmetry factor for the ``fall-time'' for $10^{20}$
eV iron and proton respectively.
The distribution is quite symmetric when plotted as a function
of the logarithm of $\sec\theta$.
To determine the position of this peak we fit a
normal distribution in the logarithm of $\sec\theta$.\\
\begin{figure}
\begin{minipage}[t]{7.5cm}
\begin{center}
\includegraphics*[width=1.1\textwidth,angle=0,clip]{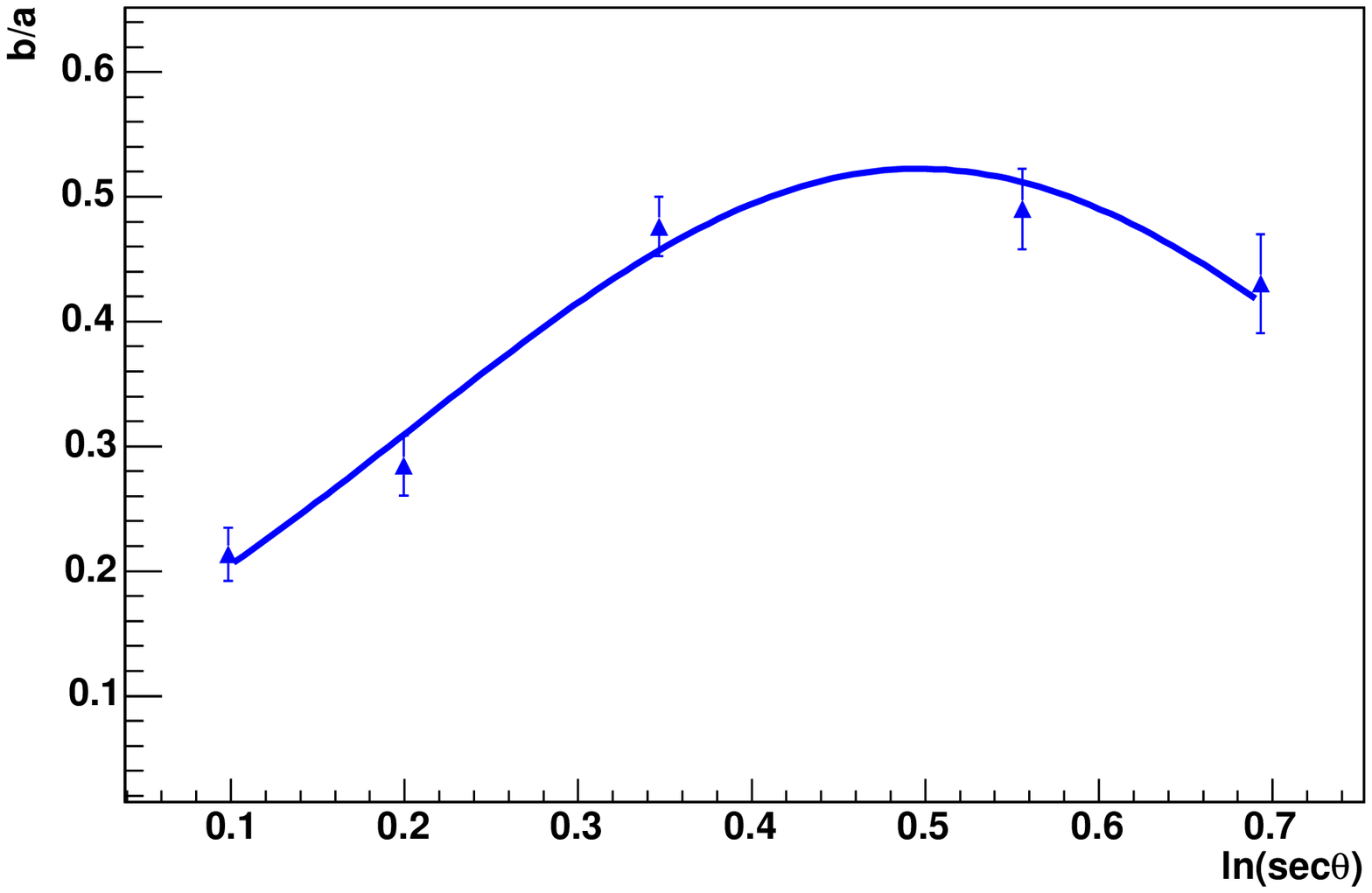}
\end{center}
\caption{Asymmetry factor as a function of $\ln \sec(\theta)$
corresponding to  ``fall-time'', for 100 EeV iron.}
\label{secmaxfall}
\end{minipage}
\hfill
\begin{minipage}[t]{7.5cm}
\begin{center}
\includegraphics*[width=1.1\textwidth,angle=0,clip]{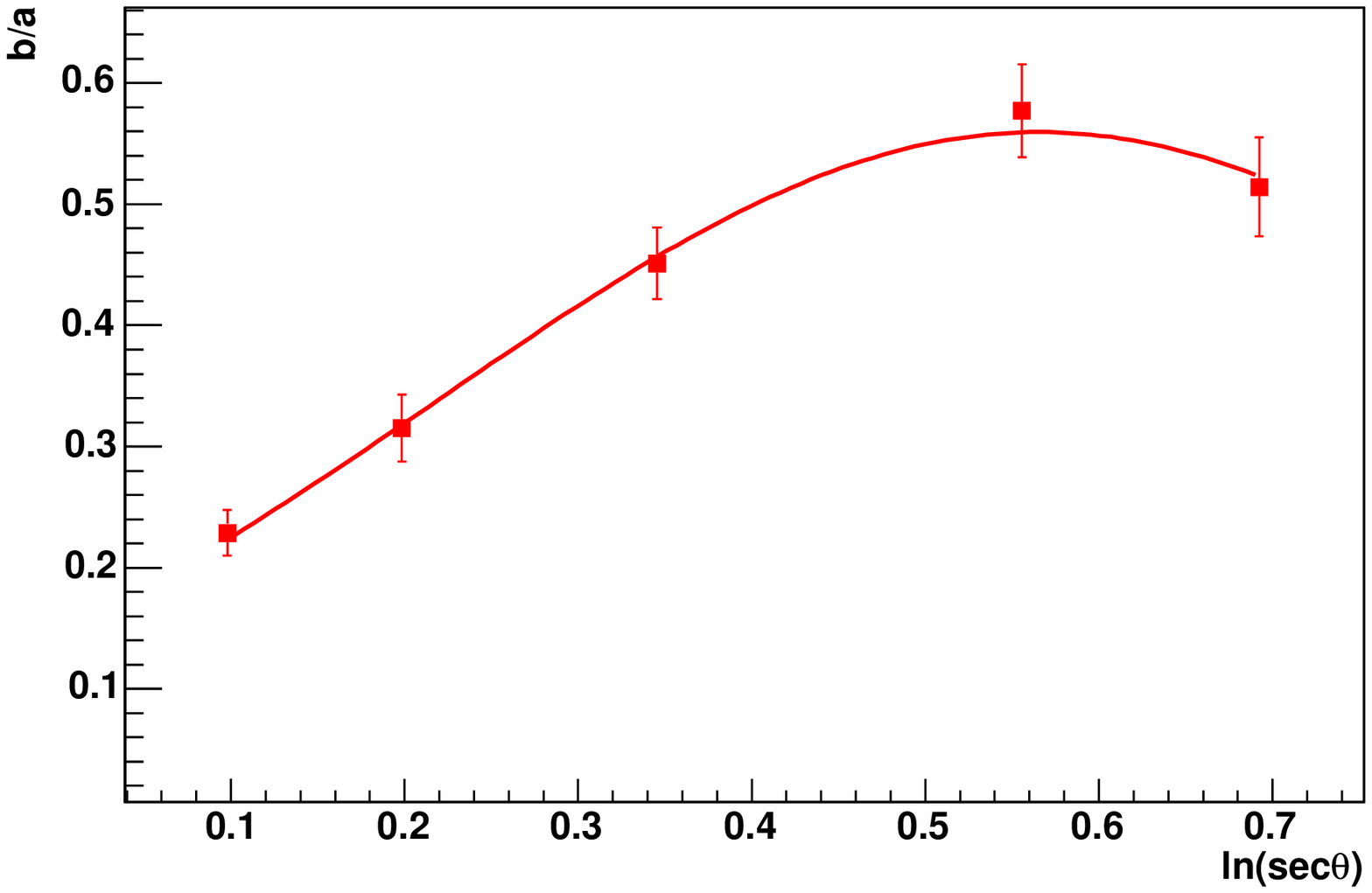}
\end{center}
\caption{Asymmetry factor as a function of $\ln \sec(\theta)$
corresponding to  ``fall-time'', for 100 EeV proton}
\label{secmaxfallproton}
\end{minipage}
\hfill
\end{figure}
The location of the peak, $\ln (\sec \theta)_{max}$,  as a function
of primary energy is shown in Figure ~\ref{asymmaxfall} for proton and iron primaries 
for the ``fall-time''. 
%In Figure 
%~\ref{asimmaxrise} we present $\ln (\sec \theta)_{max}$ for the ``rise-time''.
Clearly, 
%The relationship between $\ln (\sec \theta)_{max}$ and $X_{asymax}$ is:
$\ln(\sec \theta)_{max}=\ln(X_{asymax}/t)$. The error bars are the errors from the fit.
The position of the peak
in the dependence of the asymmetry factor with $sec\theta$ is
different for different primaries and 
allows then to help separating primary species.
The ``fall-time'' seems to be a better discriminating parameter than the ``rise-time''
at all primary energies. It is worth mentioning that the value of the asymmetry at the 
peak is also sensitive, although to a less extent, to the primary
composition.\\
%In Figures ~\ref{asymmaxfall} and ~\ref{asymmaxrise} we %show the maximum value of the asym%metry factor obtained from 
%the fit for proton ({\tiny $\square$}) and iron ($\blacktriangle$) initiated 
%showers as a function of primary energy
%corresponding to  ``fall-time'' and ``rise-time'' respectively.
We also found  that the difference in the peak position
for proton and iron expressed in g. cm$^{-2}$ is of the same
order as the difference in $X_{max}$ position for each
energy bin. 
This is not surprising since observables related to
composition are in some way a measurement of the stage of shower development
and then, directly correlated with the shower maximum $X_{max}$.\\
\begin{figure}
\begin{minipage}[t]{15cm}
\begin{center}
\includegraphics*[width=0.85\textwidth,angle=0,clip]{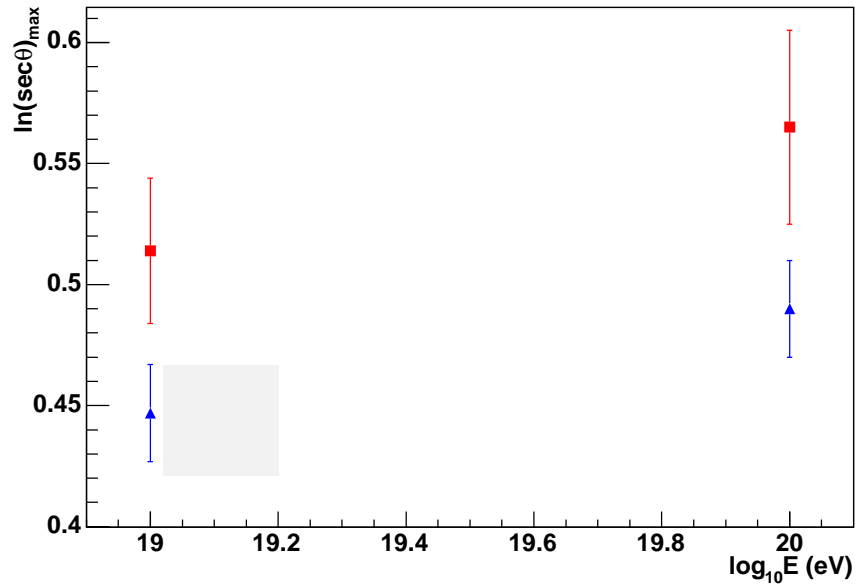}
\end{center}
\caption{Maximum value of the asymmetry factor obtained from the the fit for proton ({\tiny $\square$}) and iron ($\blacktriangle$) initiated showers as a function of primary energy
corresponding to  ``fall-time''.}
\label{asymmaxfall}
\end{minipage}
%\hfill
%\begin{minipage}[t]{7.5cm}
%\begin{center}
%\includegraphics*[width=1.0\textwidth,angle=0,clip]{XAsymmaxRise.eps}
%\end{center}
%\caption{Maximum value of the asymmetry factor obtained from the the fit for proton ({\tiny $\square$}) and iron ($\blacktriangle$) initiated showers as a function of primary energy
%corresponding to  ``rise-time''.}
%\label{asymmaxrise}
%\end{minipage}
%\hfill
\end{figure}
%The new parameter introduced in this work, the position of the peak
%in the distribution of  ``rise-time'' or ``fall-time'' with $sec\theta$ ( or
%the atmospheric depth), is very sensitive to the primary mass and then, %provides
%an additional variable to separate showers initiated by heavy and light primaries.
%Certainly, using a large set of inclined showers, the information
%of the mean mass can be extracted statistically from
%the data sample.

\section{Conclusions}

Data collected by the surface detector of the Pierre
Auger Observatory have proven to be extremely rich
for the inference of the characteristic of the primary particle.
%We have performed a study of the 
%detector signals using Monte Carlo, to explore new variables related to composition
%for inclined showers.  
In particular, the observed azimuth angle asymmetries in
time distribution of the signals 
in showers with zenith angles lower than $70^{\circ}$,
which is a unique feature of the Pierre Auger Observatory. \\
We focussed our analysis on the study of the azimuthal asymmetry of both the ``rise-time'' and
``fall-time'' that are sensitive to the presence of the electromagnetic and muonic components
of the shower and, because of this to primary composition.
From a study of the dependence of the azimuthal asymmetries in these timing variables
with the depth in the atmosphere
using Monte Carlo simulations for proton and iron showers in different
energy ranges, we have found a new observable for primary mass discrimination:
A clear peak appears in the distribution of the asymmetry factor with atmospheric depth,
the position of which is sensitive to primary mass. 
With a high statistics sample of inclined showers we expect to be able to 
obtain information on the primary composition with good precision
using the method described in this work.
An important feature of this study is the weak dependence of
the new observable with energy, requiring only the knowledge of 
the core position with good resolution.\\
We have applied the method to signal pulse shape parameters but it could also
be applied to some other pulse shape parameters or
shower observables like the shower age.

\end{document}